\documentclass[fleqn,twoside]{article}
\usepackage{espcrc2,epsf,epsfig,psfrag,graphicx}
\usepackage{graphicx}
\usepackage[figuresright]{rotating}
\newcommand{\beq}{\begin{equation}}
\newcommand{\eeq}{\end{equation}}
\newcommand{\bea}{\begin{eqnarray}}
\newcommand{\eea}{\end{eqnarray}}
\def\simlt{\buildrel < \over {_{\sim}}}
\def\simgt{\buildrel > \over {_{\sim}}}

\newcommand{\AmS}{{\protect\the\textfont2
  A\kern-.1667em\lower.5ex\hbox{M}\kern-.125emS}}

\title{Neutrino Mass Models}

\author{S. F. King\address[MCSD]{Department of Physics and Astronomy, \\ 
        University of Southampton, Southampton SO17 1BJ, U.K.}%
        \thanks{PPARC Senior Research Fellow.}}
       
\begin{document}

\begin{abstract}
I review promising approaches to neutrino mass models,
focussing on three neutrino
patterns of neutrino masses and mixing angles,
and the corresponding Majorana mass matrices.
I discuss the see-saw mechanism, and show how it may be applied
in a very natural way to give a neutrino mass hierarchy with
large atmospheric and solar angles by assuming 
single right-handed neutrino dominance. 
A theoretical framework for understanding quark and lepton
(including neutrino) masses and mixing angles
based on SUSY, GUTs and Family symmetry is then described,
and sample models which involve single right-handed neutrino
dominance are discussed.
\vspace{1pc}
\end{abstract}

\maketitle

\section{INTRODUCTION}

Recent SNO results \cite{Hallin} when combined
with other solar neutrino data especially that of Super-Kamiokande
\cite{Smy} strongly favour the large mixing angle (LMA) MSW solar solution 
\cite{Smirnov,Lisi,Valle} with three active light neutrino states, and
$\theta_{12} \approx \pi/6$, 
$\Delta m_{21}^2\approx 5\times 10^{-5}{\rm eV}^2$.
The atmospheric neutrino data is consistent with
maximal $\nu_{\mu}- \nu_{\tau}$ neutrino mixing \cite{Shiozawa}
$\theta_{23} \approx \pi/4$
with $|\Delta m_{32}^2|\approx 2.5\times 10^{-3}{\rm eV}^2$
and the sign of $\Delta m_{32}^2$ undetermined. 
The CHOOZ experiment limits $\theta_{13} \simlt 0.2$
over the favoured atmospheric range \cite{Apollonio:1999ae}.

In this talk I concentrated on promising approaches rather than
attempting a model survey \cite{Altarelli:2002hx}.
Unfortunately I did not have time
to discuss alternative approaches based on large extra
dimensions or R-parity violating supersymmetry.
Similarly I only considered three active neutrinos.
If the LSND signal \cite{Drexlin} is confirmed by
MiniBoone \cite{Tayloe} then this may herald an era of non-standard
neutrino physics. Fortunately such ideas were fully discussed in
\cite{Valle}. 

In the remainder I introduce neutrino masses and mixing
angles, show how to construct the MNS matrix,
classify Majorana mass matrices, review the see-saw mechanism, 
explain single right-handed neutrino dominance, motivate
SUSY, GUTs and Family symmetry approaches and discuss models
which involve all these ideas.

\section{NEUTRINO MASSES AND MIXING ANGLES}

The minimal neutrino sector required to account for the
atmospheric and solar neutrino oscillation data consists of
three light physical neutrinos with left-handed flavour eigenstates,
$\nu_e$, $\nu_\mu$, and $\nu_\tau$, defined to be those states
that share the same electroweak doublet as the left-handed
charged lepton mass eigenstates.
Within the framework of three--neutrino oscillations,
the neutrino flavor eigenstates $\nu_e$, $\nu_\mu$, and $\nu_\tau$ are
related to the neutrino mass eigenstates $\nu_1$, $\nu_2$, and $\nu_3$
with mass $m_1$, $m_2$, and $m_3$, respectively, by a $3\times3$ 
unitary matrix called the MNS matrix $U_{MNS}$
\cite{Maki:1962mu,Lee:1977qz}
\begin{equation}
\left(\begin{array}{c} \nu_e \\ \nu_\mu \\ \nu_\tau \end{array} \\ \right)=
\left(\begin{array}{ccc}
U_{e1} & U_{e2} & U_{e3} \\
U_{\mu1} & U_{\mu2} & U_{\mu3} \\
U_{\tau1} & U_{\tau2} & U_{\tau3} \\
\end{array}\right)
\left(\begin{array}{c} \nu_1 \\ \nu_2 \\ \nu_3 \end{array} \\ \right)
\; .
\end{equation}

Assuming the light neutrinos are Majorana,
$U_{MNS}$ can be parameterized in terms of three mixing angles
$\theta_{ij}$ and three complex phases $\delta_{ij}$.
A unitary matrix has six phases but three of them are removed 
by the phase symmetry of the charged lepton Dirac masses.
Since the neutrino masses are Majorana there is no additional
phase symmetry associated with them, unlike the case of quark
mixing where a further two phases may be removed.
The MNS matrix may be parametrised by 
a product of three complex Euler rotations,
\begin{equation}
U_{MNS}=U_{23}U_{13}U_{12}
\end{equation}
where
\begin{equation}
U_{23}=
\left(\begin{array}{ccc}
1 & 0 & 0 \\
0 & c_{23} & s_{23}e^{-i\delta_{23}} \\
0 & -s_{23}e^{i\delta_{23}} & c_{23} \\
\end{array}\right)
\end{equation}

\begin{equation}
U_{13}=
\left(\begin{array}{ccc}
c_{13} & 0 & s_{13}e^{-i\delta_{13}} \\
0 & 1 & 0 \\
-s_{13}e^{i\delta_{13}} & 0 & c_{13} \\
\end{array}\right)
\end{equation}

\begin{equation}
U_{12}=
\left(\begin{array}{ccc}
c_{12} & s_{12}e^{-i\delta_{12}} & 0 \\
-s_{12}e^{i\delta_{12}} & c_{12} & 0\\
0 & 0 & 1 \\
\end{array}\right)
\end{equation}

where $c_{ij} = \cos\theta_{ij}$ and $s_{ij} = \sin\theta_{ij}$.
Note that the allowed range of the angles is
$0\leq \theta_{ij} \leq \pi/2$. 
Since we have assumed that the neutrinos are Majorana, 
there are two extra phases, but only one combination
$\delta = \delta_{13}-\delta_{23}-\delta_{12}$
affects oscillations.

\begin{figure}[t] 
    \includegraphics[width=0.46\textwidth]{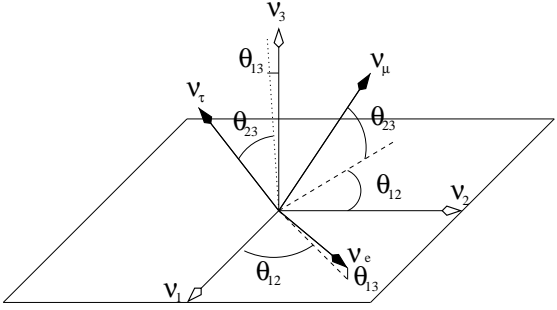}
\vspace*{-4mm}
    \caption{A graphical illustration of the MNS mixing angles.
This figure ignores the phases
so that the MNS matrix is constructed
as a product of three Euler rotations $U_{MNS}=R_{23}R_{13}R_{12}$.
The atmospheric angle is $\theta_{23} \approx \pi/4$, 
the CHOOZ angle is $\theta_{13} \simlt 0.2$, and the solar angle is
$\theta_{12} \approx \pi/6$. }   
\label{angles}
\vspace*{-2mm}
\end{figure}

Ignoring phases, the relation between 
the neutrino flavor eigenstates $\nu_e$, $\nu_\mu$, and $\nu_\tau$ and
the neutrino mass eigenstates $\nu_1$, $\nu_2$, and $\nu_3$
is just given as a product of three Euler rotations as
depicted in Fig.\ref{angles}.

\begin{figure}[t] 
\includegraphics[width=0.46\textwidth]{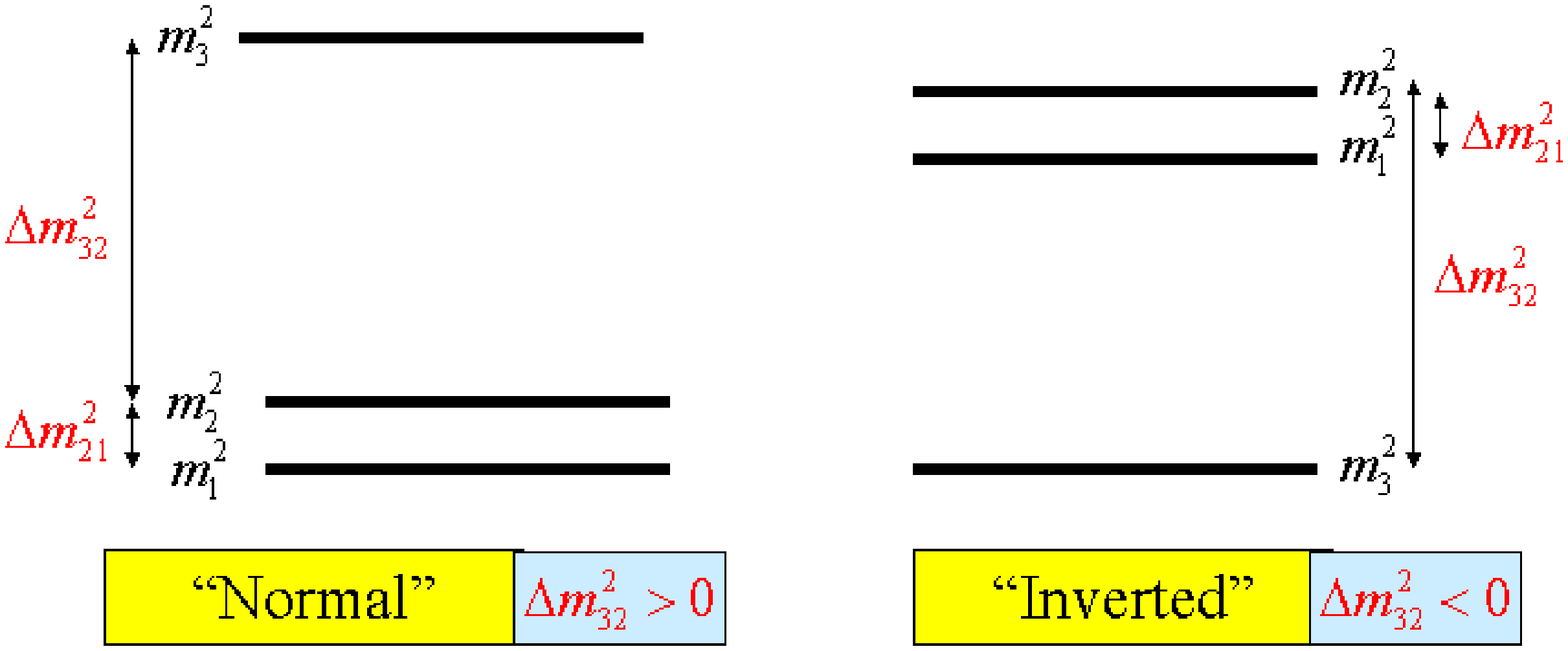}
\vspace*{-4mm}
    \caption{Alternative neutrino mass 
patterns that are consistent with neutrino
oscillation explanations of the atmospheric and solar data.
The absolute scale of neutrino masses is not
fixed by oscillation data
and the lightest neutrino mass may vary from 0.0-0.6 eV.}   
\label{fig1}
\vspace*{-2mm}
\end{figure}

There are basically two
patterns of neutrino mass squared orderings 
consistent with the atmospheric and solar data as shown in 
Fig.\ref{fig1}.

It is clear that neutrino oscillations, which 
only depend on $\Delta m_{ij}^2\equiv m_i^2-m_j^2$, 
gives no information about the absolute value of the neutrino mass squared
eigenvalues $m_i^2$ in Fig\ref{fig1}.
Recent results from the 2df galaxy redshift survey
indicate that $\sum m_i <1.8 {\rm eV} (95\%C.L.)$
under certain mild assumptions \cite{Elgaroy:2002bi,Hannestad}.
Combined with the solar and atmospheric oscillation data
this brackets the heaviest neutrino mass to be
in the approximate range 0.04-0.6 eV. The fact that the mass of the
heaviest neutrino is known to within an order of magnitude represents
remarkable progress in neutrino physics over recent years.

\section{CONSTRUCTING THE MNS MATRIX}

From a model building perspective the neutrino 
and charged lepton masses are given by the eigenvalues
of a complex charged lepton mass matrix
$m^E_{LR}$ and a complex symmetric neutrino Majorana matrix
$m_{LL}$, obtained by diagonalising these mass
matrices,
\beq
V^{E_L}m^E_{LR}{V^{E_R}}^{\dagger}=
\left( \begin{array}{ccc}
m_e & 0 & 0    \\
0 & m_{\mu} & 0 \\
0 & 0 & m_{\tau}
\end{array}
\right) 
\label{diag1}
\eeq
\beq
V^{\nu_L}m_{LL}{V^{\nu_L}}^T=
\left( \begin{array}{ccc}
m_1 & 0 & 0    \\
0 & m_2 & 0 \\
0 & 0 & m_3
\end{array}
\right) 
\label{diag2}
\eeq
where $V^{E_L}$, $V^{E_R}$, $V^{\nu_L}$ are unitary tranformations 
on the left-handed charged lepton fields $E_L$, right-handed charged
lepton fields $E_R$, and left-handed neutrino fields $\nu_L$
which put the mass matrices into diagonal form with real
eigenvalues. 

The MNS matrix is then constructed by
\beq
U_{MNS}=V^{E_L}{V^{\nu_L}}^{\dagger}
\label{MNS}
\eeq

The MNS matrix is constructed in Eq.\ref{MNS} as a product
of a unitary matrix from the charged lepton sector $V^{E_L}$
and a unitary matrix from the neutrino sector ${V^{\nu_L}}^{\dagger}$.
Each of these unitary matrices may be parametrised by its own
mixing angles and phases analagous to the MNS parameters.
As shown in \cite{King:2002nf} the MNS matrix can be expanded
in terms of neutrino and charged lepton mixing angles and phases
to leading order in the charged lepton mixing angles which 
are assumed to be small,
\bea
s_{23}e^{-i\delta_{23}}
& \approx &
s_{23}^{\nu_L}e^{-i\delta_{23}^{\nu_L}}
-\theta_{23}^{E_L} 
c_{23}^{\nu_L}e^{-i\delta_{23}^{E_L}}
\label{chlep23}
\\
\theta_{13}e^{-i\delta_{13}}
& \approx &
\theta_{13}^{\nu_L}e^{-i\delta_{13}^{\nu_L}}
-\theta_{13}^{E_L}c_{23}^{\nu_L}e^{-i\delta_{13}^{E_L}} \nonumber \\
& + &
\theta_{12}^{E_L}s_{23}^{\nu_L}e^{i(-\delta_{23}^{\nu_L}-\delta_{12}^{E_L})}
\label{chlep13}
\\
s_{12}e^{-i\delta_{12}} 
& \approx &
s_{12}^{\nu_L}e^{-i\delta_{12}^{\nu_L}}
+\theta_{23}^{E_L}s_{12}^{\nu_L}e^{-i\delta_{12}^{\nu_L}}
\nonumber \\
& + & \theta_{13}^{E_L}
c_{12}^{\nu_L}s_{23}^{\nu_L}e^{i(\delta_{23}^{\nu_L}-\delta_{13}^{E_L})}
\nonumber \\
& - & \theta_{12}^{E_L}
c_{23}^{\nu_L}c_{12}^{\nu_L}e^{-i\delta_{12}^{E_L}}
\label{chlep12}
\eea
Clearly $\theta_{13}$
receives important contributions not just from $\theta_{13}^{\nu_L}$,
but also from the charged lepton angles
$\theta_{12}^{E_L}$, and $\theta_{13}^{E_L}$.
In models where $\theta_{13}^{\nu_L}$ is
extremely small, $\theta_{13}$ may originate almost entirely from 
the charged lepton sector.
Charged lepton contributions could also be important in models
where $\theta_{12}^{\nu_L}=\pi /4$, since charged lepton mixing angles
may allow consistency with the LMA MSW solution.
Such effects are important for the inverted hierarchy model
\cite{King:2001ce,King:2002nf}.

\section{NEUTRINO MAJORANA MASS MATRICES}

For many (but not all) purposes it is convenient to forget about the 
division between charged lepton and neutrino mixing angles and 
work in a basis where the charged lepton mass matrix is diagonal.
Then the MNS angles and phases simply correspond to the neutrino ones.
In this special basis the mass matrix is given from Eq.\ref{diag2} and
Eq.\ref{MNS} as
\beq
m_{LL}=
U_{MNS}
\left( \begin{array}{ccc}
m_1 & 0 & 0    \\
0 & m_2 & 0 \\
0 & 0 & m_3
\end{array}
\right) 
U_{MNS}^T
\label{mLLnu}
\eeq
For a given assumed form of $U_{MNS}$ and set of neutrino masses 
$m_i$ one may use Eq.\ref{mLLnu} to ``derive'' the form of the neutrino 
mass matrix $m_{LL}$, and this results in the candidate 
mass matrices in Table \ref{table1} \cite{Barbieri:1998mq}.

In Table \ref{table1} the mass matrices are classified into two types:

Type I - small neutrinoless double beta decay

Type II - large neutrinoless double beta decay

They are also classified into the limiting cases consistent with the 
mass squared orderings in Fig.\ref{fig1}:

A - Normal hierarchy $m_1^2,m_2^2\ll m_3^2$

B - Inverted hierarchy $m_1^2 \approx m_2^2 \gg m_3^2$

C - Approximate degeneracy $m_1^2\approx m_2^2\approx m_3^2$

Thus according to our classification there is only one neutrino
mass matrix consistent with the normal neutrino mass hierarchy which 
we call Type IA, corresponding to the leading order neutrino masses
of the form $m_i=(0,0,m)$. For the inverted hierarchy there are two
cases, Type IB corresponding to $m_i=(m,-m,0)$ or Type IIB
corresponding to $m_i=(m,m,0)$. For the approximate degeneracy cases
there are three cases, Type IC correponding to $m_i=(m,-m,m)$
and two examples of Type IIC corresponding to either 
$m_i=(m,m,m)$ or $m_i=(m,m,-m)$. 

At present experiment allows any of the matrices in Table
\ref{table1}. In future it will be possible to uniquely specify the
neutrino matrix in the following way:

1. Neutrinoless double beta effectively measures the 11 element 
of the mass matrix $m_{LL}$ corresponding to  
\beq
\beta \beta_{0\nu}\equiv \sum_iU_{ei}^2m_i
\eeq
and is clearly capable of resolving Type I from Type II cases
according to the bounds given in Table \ref{table1} \cite{Pascoli:2002xq}.
There has been a recent claim of a signal in neutrinoless double 
beta decay correponding to $\beta \beta_{0\nu}=0.11-0.56$ eV at 95\% C.L.
\cite{Klapdor-Kleingrothaus:2001ke}.
However this claim has been criticised by two groups
\cite{Feruglio:2002af}, \cite{Aalseth:2002dt} and in turn this
criticism has been refuted \cite{Klapdor-Kleingrothaus:2002kf}. 
Since the Heidelberg-Moscow experiment has almost reached its full
sensitivity, we may have to wait for a next generation experiment
such as GENIUS \cite{genius} which is capable 
of pushing down the sensitivity to 0.01 eV to resolve this question.

2. A neutrino factory will measure the sign of $\Delta m_{32}^2$
and resolve A from B.

3. Tritium beta decay experiments are sensitive to C
since they
measure the ``electron neutrino mass'' defined by
\beq
|m_{\nu_e}|\equiv \sum_i|U_{ei}|^2|m_i|.
\eeq
For example the KATRIN \cite{katrin} experiment has a proposed sensitivity of 
0.35 eV. As already mentioned the galaxy power spectrum combined with 
solar and atmospheric oscillation data already limits the degenerate
neutrino mass to be less than about 0.6 eV, and this limit is also
expected to improve in the future. Also it is worth mentioning that
in future it may be possible to measure neutrino masses from 
gamma ray bursts using time of flight techniques in principle down to
0.001 eV \cite{Choubey:2002bh}.

Type IIB and C involve small fractional 
mass splittings $|\Delta m_{ij}^2| \ll m^2$ which are unstable 
under radiative corrections \cite{Ellis:1999my}, and even the most
natural Type IC case is difficult to implement 
\cite{Barbieri:1999km},\cite{Chankowski:2001fp}.
Types IA and IB seem to be the most natural
and later we shall focus on the normal hierarchy Type IA, 
\beq
m_{LL} \sim \left(
\begin{array}{ccc}
0 & 0 & 0 \\
0 & 1 & 1\\
0 & 1 & 1 \\
\end{array}
\right)\frac{m}{2}
\label{hier}
\eeq
However even Type IA models appear to have some remaining naturalness
problem since  
$m_3 \sim \sqrt{|\Delta m_{32}^2|}\sim 5.10^{-2}$ eV and 
$m_2 \sim \sqrt{|\Delta m_{21}^2|}\sim 7.10^{-3}$ eV, 
compared to the natural expectation $m_2 \sim m_3$.
The question may be phrased
in technical terms as one of understanding why 
the sub-determinant of the mass matrix in Eq.\ref{hier} is small:
\beq
det \left(
\begin{array}{cc}
m_{22} & m_{23}\\
m_{23} & m_{33} \\
\end{array}
\right)\ll m^2.
\label{det}
\eeq

\section{THE SEE-SAW MASS MECHANISM}

Before discussing the see-saw mechanism it is worth first reviewing
the different types of neutrino mass that are possible. So far we
have been assuming that neutrino masses are Majorana masses of the form
\beq
m_{LL}\overline{\nu_L}\nu_L^c
\label{mLL}
\eeq
where $\nu_L$ is a left-handed neutrino field and $\nu_L^c$ is
the CP conjugate of a left-handed neutrino field, in other words
a right-handed antineutrino field. Such Majorana masses are possible
to qsince both the neutrino and the antineutrino
are electrically neutral and so 
Majorana masses are not forbidden by electric charge conservation.
For this reason a Majorana mass for the electron would
be strictly forbidden. Majorana neutrino masses ``only''
violate lepton number conservation. 
If we introduce right-handed neutrino fields then there are two sorts
of additional neutrino mass terms that are possible. There are
additional Majorana masses of the form
\beq
M_{RR}\overline{\nu_R}\nu_R^c
\label{MRR}
\eeq
where $\nu_R$ is a right-handed neutrino field and $\nu_R^c$ is
the CP conjugate of a right-handed neutrino field, in other words
a left-handed antineutrino field. In addition there are
Dirac masses of the form
\beq
m_{LR}\overline{\nu_L}\nu_R.
\label{mLR}
\eeq
Such Dirac mass terms conserve lepton number, and are not forbidden 
by electric charge conservation even for the charged leptons and
quarks. 

In the Standard Model Dirac mass terms for charged leptons and quarks
are generated from Yukawa couplings
to a Higgs doublet whose vacuum expectation value gives the Dirac
mass term. Neutrino masses are zero in the Standard Model because
right-handed neutrinos are not present, and also 
because the Majorana mass terms in Eq.\ref{mLL}
require Higgs triplets in order to be generated at the renormalisable
level (although non-renormalisable operators can be written down
\cite{Valle}). Higgs triplets are phenomenologically
disfavoured so the simplest way to generate neutrino masses
from a renormalisable theory is to introduce right-handed neutrinos.
Once this is done then the types of neutrino mass discussed
in Eqs.\ref{MRR},\ref{mLR} (but not Eq.\ref{mLL} since we 
have not introduced Higgs triplets)
are permitted, and we have the mass matrix
\begin{equation}
\left(\begin{array}{cc} \overline{\nu_L} & \overline{\nu^c_R}
\end{array} \\ \right)
\left(\begin{array}{cc}
0 & m_{LR}\\
m_{LR}^T & M_{RR} \\
\end{array}\right)
\left(\begin{array}{c} \nu_L^c \\ \nu_R \end{array} \\ \right)
\label{matrix}
\end{equation}
Since the right-handed neutrinos are electroweak singlets
the Majorana masses of the right-handed neutrinos $M_{RR}$
may be orders of magnitude larger than the electroweak
scale. In the approximation that $M_{RR}\gg m_{LR}$ 
the matrix in Eq.\ref{matrix} may be diagonalised to 
yield effective Majorana masses of the type in Eq.\ref{mLL},
\beq
m_{LL}=m_{LR}M_{RR}^{-1}m_{LR}^T
\label{seesaw}
\eeq
This is the see-saw mechanism \cite{seesaw,seesaw2}. It not only generates
Majorana mass terms of the type $m_{LL}$, but also naturally makes them 
smaller than the Dirac mass terms by a factor of $m_{LR}/M_{RR}\ll 1$.
One can think of the heavy right-handed neutrinos as being integrated
out to give non-renormalisable Majorana operators suppressed
by the heavy mass scale $M_{RR}$.

In a realistic model with three left-handed neutrinos and
three right-handed neutrinos the Dirac masses $m_{LR}$
are a $3\times 3$ (complex) matrix and the heavy Majorana masses $M_{RR}$
form a separate $3\times 3$ (complex symmetric) matrix. 
The light effective Majorana
masses $m_{LL}$ are also a $3\times 3$ (complex symmetric) matrix and 
continue to be given from Eq.\ref{seesaw} which
is now interpreted as a matrix product. From a model building
perspective the fundamental parameters which must be input
into the see-saw mechanism are the Dirac mass matrix $m_{LR}$ and 
the heavy right-handed neutrino Majorana mass matrix $M_{RR}$.
The light effective left-handed Majorana mass
matrix $m_{LL}$ arises as an output according to the
see-saw formula in Eq.\ref{seesaw}. The goal of see-saw model
building is therefore to choose input see-saw matrices
$m_{LR}$ and $M_{RR}$ that will give rise to one of the successful
matrices $m_{LL}$ in Table \ref{table1}.

\section{SINGLE RIGHT HANDED NEUTRINO DOMINANCE}

We now show how the input see-saw matrices can be simply chosen to 
give the Type IA matrix in Eq.\ref{hier}, with the property of a
naturally small sub-determinant in Eq.\ref{det} using a mechanism
first suggested in \cite{King:1998jw}.
The idea was developed in \cite{King:1999cm} where it was called
single right-handed neutrino dominance (SRHND) . SRHND was first successfully
applied to the LMA MSW solution in \cite{King:2000mb}.  

The SRHND mechanism is most simply described 
assuming three right-handed neutrinos
in the basis where the right-handed neutrino mass matrix is diagonal
although it can also be developed in other bases 
\cite{King:1999cm,King:2000mb}. In this basis we write the input
see-saw matrices as
\begin{equation}
M_{RR}=
\left( \begin{array}{ccc}
X' & 0 & 0    \\
0 & X & 0 \\
0 & 0 & Y
\end{array}
\right) 
\label{seq1}
\end{equation}
\begin{equation}
m_{LR}=
\left( \begin{array}{ccc}
a' & a & d    \\
b' & b & e \\
c' & c & f
\end{array}
\right) 
\label{dirac}
\end{equation}
In \cite{King:1998jw} it was suggested that one of the right-handed neutrinos 
may dominante the contribution to $m_{LL}$ if it is lighter than
the other right-handed neutrinos. 
The dominance condition was subsequently generalised to 
include other cases where the right-handed neutrino may be
heavier than the other right-handed neutrinos but dominates due to its larger
Dirac mass couplings \cite{King:1999cm}.
In any case the dominant neutrino may be taken to be the third one 
without loss of generality. Assuming SRHND then Eqs.\ref{seesaw},
\ref{seq1}, \ref{dirac} give
\beq
m_{LL}
\approx
\left( \begin{array}{ccc}
\frac{d^2}{Y}
& \frac{de}{Y}
& \frac{df}{Y}    \\
.
& \frac{e^2}{Y} 
& \frac{ef}{Y}    \\
. & . & \frac{f^2}{Y} 
\end{array}
\right)
\label{one}
\eeq
If the Dirac mass couplings satisfy the condition 
$d\ll e\approx f$ \cite{King:1998jw}
then the matrix in Eq.\ref{one} resembles the Type IA matrix in
Eq.\ref{hier}, and furthermore has a naturally small sub-determinant as in
Eq.\ref{det}. The neutrino mass
spectrum consists of one neutrino with mass $m_3\approx (e^2+f^2)/Y$
and two approximately
massless neutrinos \cite{King:1998jw}. The atmospheric angle is
$\tan \theta_{23} \approx e/f$ \cite{King:1998jw}.
It was pointed out that small perturbations from
the sub-dominant right-handed neutrinos can then lead to a small
solar neutrino mass splitting \cite{King:1998jw}.

It was subsequently shown how to
account for the LMA MSW solution with a large solar angle
\cite{King:2000mb} by careful consideration of the sub-dominant contributions. 
One of the examples considered in \cite{King:2000mb} is when the 
right-handed neutrinos dominate sequentially,
\beq
\frac{|e^2|,|f^2|,|ef|}{Y}\gg
\frac{|xy|}{X} \gg
\frac{|x'y'|}{X'}
\label{srhnd}
\eeq
where $x,y\in a,b,c$ and $x',y'\in a',b',c'$.
Assuming SRHND with sequential sub-dominance as in
Eq.\ref{srhnd}, then Eqs.\ref{seesaw}, \ref{seq1}, \ref{dirac} give
\beq
m_{LL}
\approx
\left( \begin{array}{ccc}
\frac{a^2}{X}+\frac{d^2}{Y}
& \frac{ab}{X}+ \frac{de}{Y}
& \frac{ac}{X}+\frac{df}{Y}    \\
.
& \frac{b^2}{X}+\frac{e^2}{Y} 
& \frac{bc}{X}+\frac{ef}{Y}    \\
.
& .
& \frac{c^2}{X}+\frac{f^2}{Y} 
\end{array}
\right)
\eeq
where the contribution from the first right-handed neutrino may be 
neglected according to Eq.\ref{srhnd}. 
This was show to lead to a full neutrino mass hierarchy
\beq
m_1^2\ll m_2^2\ll m_3^2
\eeq
and, ignoring phases, the solar angle only depends
on the sub-dominant couplings and is given by 
$\tan \theta_{12} \approx a/(c_{23}b-s_{23}c)$ \cite{King:2000mb}.
The simple requirement for large solar angle is then $a\sim b-c$ 
\cite{King:2000mb}. Including phases the solar angle is given by 
the analagous result \cite{King:2002nf}
\beq
\tan \theta_{12}\approx 
\frac{|a|}{c_{23}|b|\cos \phi_b' -s_{23}|c|\cos \phi_c'}
\eeq
where the phases $\phi_b',\phi_c'$ are defined in \cite{King:2002nf}.
A related phase analysis which pointed out the importance of phases
for determining the solar angle was given in \cite{Lavignac:2002gf}.
One also obtains the interesting bound on the angle $\theta_{13}$,
\cite{King:2002nf,Lavignac:2002gf,Akhmedov:1999uw,Feruglio:2002af}
\beq
\theta_{13}\simgt  \frac{m_2}{m_3} \sim 0.1
\eeq
which implies that $\theta_{13}$ should be just below the CHOOZ limit
\cite{Vissani:1998xg}, and there is therfore a good chance
that this angle could be observed at MINOS or CNGS.

\section{SUSY, GUTs AND FAMILY SYMMETRIES}

\begin{figure}[t] 
\includegraphics[width=0.46\textwidth]{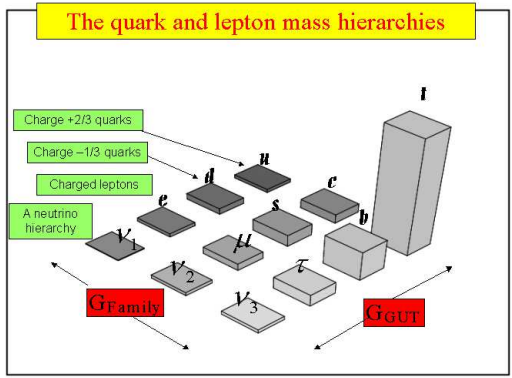}
\vspace*{-4mm}
    \caption{{\small The fermion masses are here represented by a lego plot.
We have multiplied the masses of the bottom, charm and tau by $10$, 
the strange and muon by $10^2$, the up and down
by $10^3$, the electron by $10^4$ to make the lego blocks visible.
It is natural to assume a normal neutrino 
hierarchy. We have multiplied the third neutrino mass by 
$10^{11}$ and the second neutrino mass by $10^{12}$ to make the
lego blocks visible. This underlines how incredibly light the neutrinos are.
The symmetry groups $G_{{\rm GUT}}$ and $G_{{\rm Family}}$ act
in the directions indicated.}}   
\label{fig3}
\vspace*{-2mm}
\end{figure}

One of the exciting things about the discovery of neutrino masses
and mixing angles is that this provides additional information
about the flavour problem - the problem of understanding the origin
of three families of quarks and leptons and their masses and mixing
angles (Fig.\ref{fig3}). Solutions to the flavour
problem typically involve introducing a Family symmetry 
$G_{{\rm Family}}$, 
which unifies the three families as shown in Fig.\ref{fig3}.

In the framework of the see-saw mechanism, new physics beyond the
standard model is required to violate lepton number and generate
right-handed neutrino masses which are typically around the 
GUT scale. This is also exciting since it implies that
the origin of neutrino masses is also related to some 
GUT symmetry group $G_{{\rm GUT}}$, which unifies the
fermions within each family as shown in Fig.\ref{fig3}.

It is well known that such large mass scales can only be stabilised by 
assuming a TeV scale N=1 SUSY which cancels the quadratic divergences 
of the Higgs mass, and in turn SUSY enables the gauge couplings to 
meet at the GUT scale to give a self-consistent unification picture.

Putting these ideas to together we are suggestively led to a framework of 
new physics beyond the standard model based on N=1 SUSY with commuting
GUT and Family symmetry groups,
\beq
G_{{\rm GUT}}\times G_{{\rm Family}}
\label{symmetry}
\eeq
There are many possible candidate GUT and Family symmetry groups 
some of which are listed in Table \ref{table2}. Unfortunately the
model dependence does not end there, since the details of the symmetry
breaking vacuum plays a crucial role in specifying the model and 
determining the masses and mixing angles, resulting in many models
\cite{Altarelli:2002hx,examples}. 

Another complication is that the masses and mixing angles determined
in some high energy theory must be run down to low energies using 
the renormalisation group equations (RGEs)
\cite{Babu:1993qv,Antusch:2002ek}. 
Large radiative corrections are seen \cite{Ellis:1999nk} 
when the see-saw parameters are tuned, 
since the spectrum is sensitive to small changes in the parameters.
In natural models based on SRHND the parameters are not tuned,
since the hierarchy and large atmospheric and 
solar angles arise naturally as discussed in the previous section.
Therefore in SRHND models the radiative
corrections to neutrino masses and mixing angles are only expected
to be a few per cent, and this has been verified numerically
\cite{King:2000hk}.

In the next section we give two examples of models with the symmetry structure
of Eq.\ref{symmetry} which incorporate SRHND.

\section{TWO EXAMPLES}

\subsection{$SU(5)_{{\rm GUT}}\times U(1)_{{\rm Family}}$}
A simple example assigns $U(1)_{{\rm Family}}$ charges to 
$SU(5)_{{\rm GUT}}$ multiplets labelled by a family index,
$\overline{5_i}=(L_i,D^c_i)$, ${10_i}=(Q,U^c_i,E^c_i)$, $1_i=N^c_i$,
as follows \cite{Altarelli:1998ns}:
\beq
\overline{5_i}=(3,0,0), \ \ {10_i}=(3,2,0), \ \ 1_i=(1,-1,0)
\eeq
Note that three right-handed neutrinos have been added by hand.
This results in the neutrino see-saw matrices \cite{Altarelli:1998ns}
\beq
m_{LR}\sim
\left( \begin{array}{ccc}
\lambda^4 & \lambda^2 & \lambda^3    \\
\lambda & \lambda & 1 \\
\lambda & \lambda & 1
\end{array}
\right) 
\label{mLR1}
\eeq
\beq
M_{RR}\sim
\left( \begin{array}{ccc}
\lambda^2 & 1 & \lambda    \\
1 & \lambda^2 & \lambda \\
\lambda & \lambda & 1
\end{array}
\right) 
\label{MRR1}
\eeq
where $\lambda \approx 0.22$ is assumed to be of order the Cabibbo angle.
Although the three right-handed neutrinos are roughly degenerate, 
this model has SRHND since 
the third right-handed neutrino dominates
due to its larger Dirac mass couplings.
A large atmospheric angle is given from
the large entry in the 23 position of the ``lop-sided'' neutrino mass matrix 
$\tan \theta_{23} \approx e/f \sim 1$.
The solar angle is $\tan \theta_{12}\sim a/(b-c) \sim \lambda$
which may be interpreted as either LMA or SMA MSW.
The quark mass matrices are given by
\beq
m_{LR}^D\sim
\left( \begin{array}{ccc}
\lambda^6 & \lambda^3 & \lambda^3    \\
\lambda^5 & \lambda^2 & \lambda^2 \\
\lambda^3 & 1 & 1
\end{array}
\right) 
\label{mDLR1}
\eeq
\beq
m_{LR}^U\sim
\left( \begin{array}{ccc}
\lambda^6 & \lambda^5 & \lambda^3    \\
\lambda^5 & \lambda^4 & \lambda^2 \\
\lambda^3 & \lambda^2 & 1
\end{array}
\right) 
\label{mULR1}
\eeq
The down mass matrix only has 
large unphysical right-handed mixing with small physical
left-handed quark mixing angles. 
The mass and mixing relations are:
\beq
V_{us}\sim \lambda, \ \ V_{ub}\sim \lambda^3, \ \ V_{cb}\sim \lambda^2
\eeq
\beq
m_u:m_c:m_t=\lambda^6:\lambda^4:1
\eeq
\beq
m_d:m_s:m_b=\lambda^6:\lambda^2:1
\eeq

\subsection{$SO(10)_{{\rm GUT}}\times SU(3)_{{\rm Family}}$}
In this model \cite{King:2001uz},
the three families are unified as triplets under 
$SU(3)_{{\rm Family}}$, and $16's$ under $SO(10)_{{\rm GUT}}$.
Further symmetries are assumed to ensure that
the vacuum alignment leads to a universal form of Dirac mass matrices
for the neutrinos, charged leptons and quarks \cite{King:2001uz},
\beq
m_{LR}\sim
\left( \begin{array}{ccc}
\epsilon^8 & \epsilon^3 & \epsilon^3    \\
-\epsilon^3 & \epsilon^2 & \epsilon^2 \\
-\epsilon^3 & \epsilon^2 & 1
\end{array}
\right) 
\label{mLR2}
\eeq
where $\epsilon \approx 0.05$ and the expansion parameter for the
down quarks and charged leptons is assumed to be larger
$\overline{\epsilon} \approx 0.15$. 

The neutrino see-saw matrices are \cite{King:2001uz},
\beq
m_{LR}\sim
\left( \begin{array}{ccc}
\epsilon^8 & \lambda_{\nu}\epsilon^3 & \lambda_{\nu}\epsilon^3    \\
-\lambda_{\nu}\epsilon^3 & a_{\nu}\epsilon^2 
& a_{\nu}\epsilon^2+\lambda'_{\nu}\epsilon^3  \\
-\lambda_{\nu}\epsilon^3 & a_{\nu}\epsilon^2+\lambda'_{\nu}\epsilon^3   & 1
\end{array}
\right) 
\label{mLR3}
\eeq
where $\lambda_{\nu}, \lambda'_{\nu}, a_{\nu}$ are order one parameters.
\beq
M_{RR}\sim
\left( \begin{array}{ccc}
\epsilon_{\nu}^4 & 0 & \epsilon_{\nu}^2    \\
0 & \epsilon_{\nu}^3 & \epsilon_{\nu}^3 \\
\epsilon_{\nu}^2 & \epsilon_{\nu}^3 & 1
\end{array}
\right) 
\label{MRR2}
\eeq
In this model the first right-handed neutrino dominates because 
the expansion parameter $\epsilon_{\nu}$ is very small so it is very light,  
leading to SRHND with sequential sub-dominance as in Eq.\ref{srhnd}.
The strong hierarchy tends to predict the LOW solution, although
LMA MSW is also possible \cite{King:2001uz}.
The results in section 6 apply after re-ordering the columns to make the
first right-handed neutrino the dominant one,
\begin{equation}
m_{LR}=
\left( \begin{array}{ccc}
d & a & a' \\
e & b & b' \\
f & c & c'
\end{array}
\right) 
\label{dirac2}
\end{equation}
The atmospheric angle is given from section 6 by 
$\tan \theta_{23} \approx e/f$ where the $SU(3)_{{\rm Family}}$ 
symmetry ensures
accurate equality $e=f=-\lambda_{\nu}\epsilon^3$.
The solar angle is given from section 6 by
$\tan \theta_{12} \sim a/(b-c)\sim -\lambda_{\nu}/\lambda'_{\nu}\sim 1$ 
where the $SU(3)_{{\rm Family}}$ symmetry 
leads to a cancellation of the leading order terms and a large solar
angle. The quark mass matrices are
\beq
m^U_{LR}\sim
\left( \begin{array}{ccc}
\epsilon^8 & \lambda_{U}\epsilon^3 & \lambda_{U}\epsilon^3    \\
-\lambda_{U}\epsilon^3 & \epsilon^2 
& \epsilon^2  \\
-\lambda_{U}\epsilon^3 & \epsilon^2 & 1
\end{array}
\right) 
\label{mULR2}
\eeq
\beq
m^D_{LR}\sim
\left( \begin{array}{ccc}
\overline{\epsilon}^8 & \lambda_{D}\overline{\epsilon}^3 
& \lambda_{D}\overline{\epsilon}^3    \\
-\lambda_{D}\overline{\epsilon}^3 & \overline{\epsilon}^2 
& \overline{\epsilon}^2  \\
-\lambda_{D}\overline{\epsilon}^3 & \overline{\epsilon}^2 & 1
\end{array}
\right) 
\label{mDLR2}
\eeq
where $\lambda_{U}, \lambda_{D}$ are order one parameters,
which provides a very good description of the quark masses
and mixing angles \cite{King:2001uz}.
The fact that the first right-handed neutrino dominates 
is the key which allows 
large neutrino mixing angles to arise from a universal form of mass
matrix with only small off-diagonal entries.

\section{CONCLUSIONS}

I have reviewed promising approaches to neutrino mass models,
focussing on three neutrino
patterns of neutrino masses and mixing angles,
and the corresponding Majorana mass matrices classified
in Table 1. I also reviewed the see-saw mechanism which at the 
present time looks like the most elegant explanation of small neutrino masses.
If the see-saw mechanism for understanding the neutrino
spectrum looks good, with SRHND it looks even better.
With SRHND the neutrino hierarchy appears very naturally, and
large atmospheric and solar angles are understood in terms
of simple analytic relations between Dirac mass elements.
I described an attractive 
theoretical framework for understanding quark and lepton
(including neutrino) masses and mixing angles
based on SUSY, GUTs and Family symmetry. Two
examples of such models were discussed which implement 
the see-saw mechanism using SRHND, where the relations responsible
for large atmospheric and solar angles could be understood
in terms of symmetries.

\begin{table*}[htb]
{\small
\caption{Leading order low energy neutrino Majorana mass matrices
$m_{LL}$ consistent with large atmospheric and solar mixing angles,
classified according to the rate of neutrinoless double beta decay
and the pattern of neutrino masses.}
\label{table1}
\newcommand{\m}{\hphantom{$-$}}
\newcommand{\cc}[1]{\multicolumn{1}{c}{#1}}
\renewcommand{\tabcolsep}{2pc} 
\renewcommand{\arraystretch}{1.2} 
\begin{tabular}{@{}|c|c|c|}
\hline
\hline
& Type I  & Type II \\ 
 & Small $\beta \beta_{0\nu}$ & Large $\beta \beta_{0\nu}$ \\
\hline
\hline
 & & \\ 
A & $\beta \beta_{0\nu}\simlt 0.0082$ eV & \\
Normal hierarchy  & & \\
$m_1^2,m_2^2\ll m_3^2$ & 
$\left(
\begin{array}{ccc}
0 & 0 & 0 \\
0 & 1 & 1\\
0 & 1 & 1 \\
\end{array}
\right)\frac{m}{2}$ & -- \\
& & \\
\hline
 & & \\ 
B & $\beta \beta_{0\nu}\simlt
0.0082$ eV & $\beta \beta_{0\nu}\simgt 0.0085$ eV\\
Inverted hierarchy & & \\
$m_1^2\approx m_2^2\gg m_3^2$ & 
$\left(
\begin{array}{ccc}
0 & 1 & 1 \\
1 & 0 & 0\\
1 & 0 & 0 \\
\end{array}
\right)\frac{m}{\sqrt{2}}$ & 
$\left(
\begin{array}{ccc}
1 & 0 & 0 \\
0 & \frac{1}{2} & \frac{1}{2}\\
0 & \frac{1}{2} & \frac{1}{2} \\
\end{array}
\right)m$
\\
& & \\
\hline                        
 & & \\ 
C &  &
$\beta \beta_{0\nu}\simgt 0.035$ eV \\
Approximate degeneracy & & diag(1,1,1)m\\
$m_1^2\approx m_2^2\approx m_3^2$ & 
$\left(
\begin{array}{ccc}
0 & \frac{1}{\sqrt{2}} & \frac{1}{\sqrt{2}} \\
\frac{1}{\sqrt{2}} & \frac{1}{2} & \frac{1}{2}\\
\frac{1}{\sqrt{2}} & \frac{1}{2} & \frac{1}{2} \\
\end{array}
\right)m$ & 
$\left(
\begin{array}{ccc}
1 & 0 & 0 \\
0 & 0 & 1\\
0 & 1 & 0 \\
\end{array}
\right)m$\\
& & \\
\hline
\hline                        
\end{tabular}\\[2pt]
}
\end{table*}

\begin{table*}[htb]
\caption{Some candidate GUT and Family symmetry groups.}
\label{table2}
\newcommand{\m}{\hphantom{$-$}}
\newcommand{\cc}[1]{\multicolumn{1}{c}{#1}}
\renewcommand{\tabcolsep}{2pc} 
\renewcommand{\arraystretch}{1.2} 
\begin{tabular}{@{}|c||c|}
\hline
\hline
$G_{{\rm GUT}}$ & $G_{{\rm Family}}$\\ 
\hline 
\hline 
$E_6$ & $SU(3)$ \\
$SO(10)$ & $SU(2)$  \\
$SU(5)$ & $U(1)$ \\
$SU(5)\times U(1)$  & $Z_N$\\
$SU(3)^3$  & $O(3)\times O(3)$ \\
$SU(4)\times SU(2)\times SU(2)$ & $SO(3)$ \\
$SU(3)\times SU(2)\times SU(2) \times U(1)$ & $S(3)\times S(3)$\\
$SU(3)\times SU(2) \times U(1) \times U(1)$ & $S(3)$ \\
\hline                                               
$SU(3)\times SU(2)\times U(1)$ & {\rm Nothing}\\
\hline                        
\hline 
\end{tabular}\\[2pt]
\end{table*}

\end{document}